\documentstyle[11pt,psfig,aaspp4]{article}

\begin{document}

\title{ PHYSICAL CONDITIONS, GRAIN TEMPERATURES, AND ENHANCED VERY SMALL
GRAINS IN THE BARNARD LOOP}

\author{Carl Heiles}
\affil{Astronomy Department, University of California,
    Berkeley, CA 94720-3411}

\author{L.M. Haffner, R.J. Reynolds}
\affil{Astronomy Department, University of Wisconsin,
    Madison, WI 53706}

\author{S.L. Tufte}
\affil{Department of Physics, Lewis \& Clark College, Portland, OR 97219}

\begin{abstract} We derive the radio spectral index of the Barnard Loop
(BL) from large-scale radio surveys at four frequencies and find it to
be a thermal source.  We use the radio data together with H$\alpha$ data
to determine the electron temperature in BL, the $\lambda$ Ori HII
region, and a high-latitude filament; all of these regions are somewhat
cooler than typical HII regions. 

	We perform least squares fits of the DIRBE diffuse IR
intensities to the 21-cm line and radio continuum intensities.  After
the resolution of a ``geometrical conundrum'', this allows us to derive
the electron density $n_e$; we find $n_e \approx 2.0$ cm$^{-3}$ and
pressure ${P \over k} \approx 24000$ cm$^{-3}$ K. 

	Grains within BL are warmer than in HI regions.  Trapped
L$\alpha$ accounts for the extra heating that is required.  This is a
general effect that needs to be accounted for in all analyses that
examine IR emission from H$^+$ regions.  Very small grains that emit 60
$\mu$m radiation are enhanced in BL relative to HI by a factor of 2-3,
while PAH's that emit 12 $\mu$m are probably deficient by a factor $\sim
2$.  \end{abstract}


\section{ INTRODUCTION}
\label{barnardintro}

	The Wisconsin H$\alpha$ Mapper (WHAM) Sky Survey maps the
H$\alpha$ emission from the diffuse Warm Ionized Medium with
unprecedented sensitivity and coverage (Haffner, Reynolds, and Tufte
1999). One of the brightest large regions is the Orion-Eridanus
superbubble. We have been combining data from many sources for this
region to gain a detailed understanding of physical conditions and
processes (Heiles, Haffner, and Reynolds 1999). Here we present our
results and analysis of a small portion of this region, the Barnard Loop
(BL). 

	O'Dell, York, and Henize (1967) developed a coherent physical
model based on UV, optical, and radio data.  Their UV images are
particularly interesting because the UV is dominated by scattered light
and allows derivation of dust grain properties; they are also of
historical interest, having been taken by astronauts aboard the Gemini
11 spacecraft.  They model BL as an ellipsoid in which the volume
density increases roughly as distance$^2$, and suggest that the ionized
portion of BL is surrounded by a neutral HI shell.  They discuss the
density structure as being produced by radiation pressure of light from
the central stars pushing on the grains.  Over a period $3 \times 10^6$
yr this would produce the observed structure, which should have
expansion velocity $\sim 9$ km s$^{-1}$.  We will defer discussion of
the kinematics and correspondence with this model to a later paper. 

	Here we restrict ourselves to a detailed analysis of radio,
optical, and diffuse IR emission data to determine the physical
conditions and dust grain properties in BL. 
Section~\ref{barnardradiosection} analyzes radio continuum surveys at
four different frequencies and finds BL to be a thermal (optically thin
free-free) emitter.  Section~\ref{barnardrjrsection} combines the radio
and WHAM H$\alpha$ data to derive upper limits on electron temperatures. 
Section~\ref{absolutecal} considers a discrepancy between the
radio/H$\alpha$ temperature and previous temperatures derived from the
NII and H$\alpha$ lines and derives a correction factor for the absolute
intensity scales of the radio and WHAM surveys. 
Section~\ref{individualobjects} adopts reddenings and turns the upper
limits into actual temperatures.  Section~\ref{dirbefits} combines the
DIRBE IR, 21-cm line (Hartmann and Burton 1997), and radio continuum
data to derive the BL electron density, grain temperature, and relative
population of very small grains and PAH's.  Finally,
Section~\ref{conclusions} summarizes our findings. 

\section{ RADIO EMISSION: THERMAL OR NONTHERMAL?}
\label{barnardradiosection}

	Reich (1978) provides a very useful history of radio
observations of the Barnard Loop (BL), which we will not repeat here. 
He concludes that there are significant background emission
fluctuations, and these prevent an accurate determination of the
spectral index of the Loop itself.  However, Haslam and Salter (1971),
studied the bright top portion at 85.5, 240, and 408 MHz and found
$T_B \propto \nu^{-2.05 \pm 0.30}$, consistent with a thermal spectrum. 

	In hopes of narrowing this conclusion, we examined four surveys
of radio continuum emission to determine whether BL is a
thermal or nonthermal radio emitter.  These were the 408 MHz survey of
Haslam et al (1983), the 820 MHz survey of Berkhuijsen (1972), the 1420
MHz survey of Reich and Reich (1986), and the 2326 MHz survey of Jonas,
Baart, and Nicolson (1998).  The three lower-frequency datasets are
available from {\it http://www.mpifr-bonn.mpg.de/survey.html}.  Jonas
provided us with the full-resolution version of the 2326 MHz survey. 

	We were concerned that the surveys might exhibit scale errors. 
We addressed this problem by using the $\lambda$Ori HII region as a
calibrator, assuming that it was an optically thin free-free (thermal)
source with $T_B \propto \nu^{-2.1}$.  Using this as a calibrator has
both good and bad points.  For the good, it is large compared to any of
the surveys' angular resolution and it is not so bright as to introduce
nonlinearities.  However, its low intensity makes its derived flux
susceptible to zerolevel and background intensity fluctuations, which
are definitely present, particularly in the lower frequency surveys
where nonthermal background radiation is brighter.  For each survey we
calculated the excess brightness of the HII region for four different
choices of ON and OFF regions; our final choice used only the brightest
part of the HII region for the ON and two immediately adjacent areas for
the OFFs, as shown by the three areas outlined by dashed lines in
Figure~\ref{quadradio}.  We adopted the 2326 MHz survey as a standard
and derived the factors required to correct the other surveys to
produce the thermal scaling.  For the four choices, these factors ranged
from $0.99 \rightarrow 1.98$, $0.98 \rightarrow 1.41$, and $1.34
\rightarrow 1.57$ for the 408, 820, and 1420 MHz surveys, respectively;
for our final choice illustrated in Figure~\ref{quadradio}, the factors
are $1.61 \pm 0.50$, $1.19 \pm 0.22$, and $1.45 \pm 0.08$.  The quoted
uncertainties are simply half the difference in ranges quoted above and
should be overestimates, because the differences occur primarily because
of background fluctuations and our final choice has the OFFs closest to
the ON. 

	If the intensity scales of the surveys were identical, these
factors would be unity.  The first two have large uncertainties, but
nevertheless seem to differ significantly from unity.  The 1420 MHz
survey certainly differs significantly from unity.  Reich and Reich
(1988) find that the 1420 MHz survey needs to be multiplied by 1.55 to
agree with the 408 MHz survey; our derived value of 1.45 differs from
this factor by only $6\%$. 

	We used these factors to make difference images.  First we
multiply each of the three low-frequency surveys by its factor.  Then we
multiply each survey image by the factor $\left( {\nu \over 2326 \ {\rm
MHz}}\right) ^{2.1}$, which converts all the temperatures to the 2326
MHz temperature if the emission is thermal.  Finally, we convolve the
2326 MHz image to the same angular resolution and subtract it.  The
result is a difference image in which departures from the arbitrary and
unknown zero-level offset represent nonthermal emission.  If there is
any morphological structure in the difference images that looks like the
BL, then the Loop exhibits nonthermal emission. 

	Figure~\ref{quadradio} exhibits the three scaled difference
images, together with the 2326 MHz image.  The diagonal cutoffs in the
images arise from the declination limits of the surveys.  The dotted
half-ring guides the eye to BL and the calibration rectangles illustrate
offsets $\pm 0.05$ K from each difference image.  All four images
exhibit stripes; they run along the scanning direction and represent
zerolevel errors from one scan to another in the original data.  The
three lower frequency difference images reveal artifacts and also
background intensity fluctuations.  The 1420 MHz difference image, lower
left, exhibits broad vertical stripes; these are zero level errors
because they extend all the way to North celestial pole.  The 408 MHz
image exhibits fluctuations over a broad angular scale, about $2 \over
3$ the size of the image, or about $8^\circ$; these are not as distinct
at 820 MHz and might be partly or wholly spurious. 

	The $\lambda$Ori HII region is invisible in the difference
images of Figure~\ref{quadradio}, which serves as a check on our method. 
BL is not recognizable in either the 820 or the 408 MHz difference
image.  The 408 MHz image does exhibit excess emission $\sim 0.12$ K to
upper right of center, but this feature has no morphological
relationship to BL.  Moreover, there is no obvious evidence of this
excess in the 820 MHz map, so the structure might be spurious.  BL seems
to be dark in the 1420 MHz difference image.  However, the most
noticeable dark portions lie on the dark broad stripes, so this
appearance may be spurious. 

	The calibration rectangles are easily recognizable at 820 and
1420 MHz, while BL is not.  Thus, the nonthermal low-frequency emission
from BL $\leq 0.05$ K.  At 2326 MHz, BL's bright horizontal top
portion (BL-TOP in Figure~\ref{quadoptical}) has $T_B \sim 0.35$ K and
the left vertical portion (BL-LEFT) has $T_B \sim 0.15$ K at 2326 MHz;
the limit of $\pm 0.05$ K means that $\leq 15\%$ and $\leq 33\%$ of the
emission is nonthermal at the lower frequencies.  At 1420 MHz BL might
exhibit a deficiency at this level, but this is contrary to nonthermal
emission, which has a steeper spectrum than thermal emission. 
Therefore, these visual estimates of upper limits on nonthermal emission
are quite conservative. 

	One usually derives the spectral index of a morphologically
distinct extended feature by making plots of intensities at many
positions at one frequency versus those at a different frequency and
performing a least squares fit.  Instead, for each survey we fit the
difference image $diff$ to the 2326 MHz brightness $T_{2326}$.  For
BL-TOP we obtained ${d\ diff \over dT_{2326}} = 0.58 \pm 0.25$, $0.05
\pm 0.01$, and $-0.25 \pm 0.03$ for the 408, 820, and 1420 MHz data,
respectively; for BL-LEFT we obtained $1.79 \pm 0.03$ and $-0.34 \pm
0.03$ for the 408 and 1420 data, respectively.  The significant negative
slopes for the 1420 data probably reflect the coincidence of BL with the
broad stripes, and the significant positive slope for BL-TOP at 408 is
not revealed in the difference image except as a morphologically
inconsistent bright blob.  In our opinion, these least squares fits are
meaningless because of artifacts and nonthermal emission fluctuations in
the low frequency images. 

	We conclude that BL exhibits no recognizable nonthermal
emission.  For BL-LEFT our conservative upper limit is that $\leq 33\%$
of the 408 MHz emission is nonthermal, and for BL-TOP $\leq 15\%$. 
Nonthermal brightness temperatures typically $\propto \nu^{-2.7}$, so at
2326 MHz these limits become $\leq 12\%$ and $\leq 4\%$, respectively. 

\section{ $T_{R/H\alpha}$: THE ELECTRON TEMPERATURE FROM RADIO AND
H$\alpha$ EMISSION} \label{barnardrjrsection}

	In this section we derive the electron temperature by comparing
the 2326 MHz and H$\alpha$ intensities, assuming that all of the 2326
MHz emission is thermal.  The important quantity for these emission
processes is the emission measure $EM$,

\begin{equation}
EM = n_e^2 L \ ,
\end{equation}

\noindent where $n_e$ is the {\it total} electron density.  Electrons
come from both H$^+$ and He$^+$, so $n_e = f_{He^+} n(H^+)$, where
$f_{He^+}$ accounts for the ionized He; below, we argue that $f_{He^+} =
1$. 

	Figure 1 compares the radio and optical images.  To facilitate
quantitative comparison of the radio and H$\alpha$ emission we convert
the usual intensity units of brightness temperature and Rayleighs to
identical units that are close to the emission measure, normalizing to
$T_4 = 1$ and $E(B-V)=0$:

\begin{mathletters}
\label{emboth}
\begin{equation}
\label{emradio_1}
EM_{2326} \equiv 1.80 \times 10^3 T_{B,obs} \ ,
\end{equation}

\begin{equation}
\label{emrjr_1}
EM_{H\alpha} \equiv 2.74 I_{H\alpha,obs} \ ,
\end{equation}

\end{mathletters}

\noindent where $EM$ is in cm$^{-6}$ pc, $T_4$ the electron temperature
in units of $10^4$ K, and $E(B-V)$ the reddening in magnitudes.  These
equations derive from Mezger and Henderson (1967) and Haffner et al
(1999), respectively.  Here $T_{B,obs}$ is the observed 2326 MHz
brightness temperature and $I_{H\alpha,obs}$ the observed integrated
line intensity in Rayleighs.  These differ from the true intensities for
two reasons.  One is the error in absolute calibration; we define the
intensity scale errors by

\begin{mathletters}
\label{scaleerror}
\begin{equation}
\label{radioscaleerror}
T_{B,true} = f_{2326} T_{B,obs}
\end{equation}
\begin{equation}
I_{H\alpha,true} = f_{H\alpha}I_{H\alpha,obs}
\end{equation}
\end{mathletters}

\noindent The other is angular size, which is discussed below.  With
these definitions, we have

\begin{mathletters}
\label{emmboth}
\begin{equation}
\label{emradio}
EM = T_4^{0.35} f_{2326}EM_{2326}\ ,
\end{equation}

\begin{equation}
\label{emrjr}
EM = T_4^{0.9} f_{He^+} e^{2.34 E(B-V)} f_{H\alpha} EM_{H\alpha} \ ,
\end{equation}

\end{mathletters}

\noindent Here we have used the extinction curve of Savage and Mathis
(1979).  Internal extinction, which occurs within the emitting region
itself, is less effective than external extinction because a significant
fraction of the extinction is scattering, and internal extinction does
not scatter photons away from the observer (Mathis 1983).  We have
assumed that the extinction occurs in a foreground cloud sufficiently
far from the emitting region that scattered H$\alpha$ photons are lost,
so that the standard extinction law applies.  If our assumption is
invalid, then reddening has less effect than above.  Combining
equations~\ref{emmboth} we have

\begin{equation} \label{t4eqn} T_{4,R/H\alpha} = \left( {f_{2326} EM_{2326}
\over f_{He^+} f_{H\alpha} EM_{H\alpha}}\right)^{1.82} e^{-4.25 E(B-V)}
\ .  \end{equation}

\noindent Here the subscript $R/H\alpha$ means that the temperature is
determined by combining radio and H$\alpha$ data. 

	Figure~\ref{quadoptical} exhibits two sets of the 2326 MHz radio
and H$\alpha$ optical images side-by-side; one set exhibits BL and the
other set a weak filament.  We delineate four regions with dashed lines:
the $\lambda$Ori HII region; BL-TOP, the top bright portion of BL;
BL-LEFT, the left dim portion of BL; and FILAMENT, a weak filament lying
near $b = -40^\circ$.  For each region we make a least squares fit for
coefficients $A$ and $B$ in the equation

\begin{equation}
\label{radoptlsfit}
EM_{2326} = A + B \cdot EM_{H\alpha} \ .
\end{equation}

	Table~\ref{barnardtable} presents the results and
Figure~\ref{halfa_radioplot} exhibits plots of $EM_{2326}$ {\it vs}
$EM_{H\alpha}$ for the four regions.  Table~\ref{barnardtable} also
presents the deduced temperatures $T_{4,R/H\alpha}$ after multiplying
the slopes $B$ for the BL and FILAMENT entries by $f_S = 1.032$ as
described immediately below.  It is not appropriate to multiply $A$ by
$f_S$ because A is an offset whose value is determined by the
background, whose angular scale exceeds $7^\circ$.

	Equations~\ref{scaleerror} are not quite correct because of
inadequate angular resolution, otherwise known as ``beam dilution''. 
The optical data integrate over a $1^\circ$-diameter circle and are
spaced by about $1^\circ$; BL and the $\lambda$Ori HII region are larger
than $1^\circ$ in all dimensions, so no correction is required. 
However, for objects $< 7^\circ$ diameter the 2326 MHz intensities must
be corrected upwards by the factor

\begin{equation}
\label{etacorr}
f_S = { \eta_{\Omega_7} \over \eta_{\Omega_S}} \ , 
\end{equation}

\noindent where $\eta$ is the beam efficiency, $\Omega_S$ the source
solid angle, and $\Omega_7$ the solid angle of a $7^\circ$-diameter
circle (Jonas et al 1998).  The $\lambda$Ori HII region has diameter
$\gtrsim 7^\circ$ so there is no correction.  Using the high-resolution
H$\alpha$ image of Isobe (1973), we model BL as a rectangle of
dimensions $1.3^\circ \times 7^\circ$, and we take $f_S = \left( {
{\eta_{\Omega_{7}} \over \eta_{\Omega_{1.3}} }} \right)^{0.5} = 1.032$,
where $\Omega_{1.3}$ is the solid angle of a $1.3^\circ$ diameter
circle. 

\section{$T_{R/H\alpha}$: CORRECTION FOR ERRORS IN ABSOLUTE CALIBRATION}

\label{absolutecal}

	The accuracy of $T_{R/H\alpha}$ depends on the accuracy of the
absolute calibrations of the 2326 MHz and H$\alpha$ datasets. Absolute
calibrations are notoriously difficult so we cannot rely on them.
Rather, we use other temperature determinations to adjust our results.

	Here we consider other temperatures determined by two
techniques, both of which compare the [NII] $\lambda 6583$ and H$\alpha$
lines.  One uses the line intensity ratio and one the difference in line
width; we denote these temperatures by $T_{NII/H\alpha}$ and
$T_{NIIwid/H\alpha wid}$, respectively.  Comparing these lines has the
great virtue that they are measured with the same instrument, so
systematic errors cancel. 

\subsection{A Two-Temperature Toy Model} \label{toymodel}

	The only significant problem from comparing the [NII] and
H$\alpha$ lines is possible temperature variations along the line of
sight.  The [NII] line emissivity increases exponentially with
temperature, while the H$\alpha$ and radio emissivities decrease with
temperature; thus, the [NII] line preferentially samples high
temperature regions.  In a low-density HII region, the equilibrium
temperature is independent of density.  However, the temperature depends
on distance from the star, becoming highest near the edge where the
spectrum of ionizing radiation hardens.  For example, models 3 and 4 by
Rubin (1968) are not too dissimilar from the $\lambda$ Ori situation and
exhibit variations $T_4 \sim 0.45 \rightarrow 0.80$.  We can get a rough
idea of these effects by considering a two-temperature toy model, $T_4 =
0.45$ and 0.80, with equal emission measures at each temperature.  We
show the results in Table~\ref{tmodels}. 

	For all entries in Table~\ref{tmodels}, $T_{4,NII/H\alpha} =
0.630$ and $T_{4,R/H\alpha} = 0.570$.  As expected, $T_{NII/H\alpha} >
T_{R/H\alpha}$, but by a modest amount compared to the difference
between the physical temperatures of the two regions.  In contrast,
$T_{NIIwid/H\alpha wid}$ depends very sensitively on the relative
turbulent velocities in the two regions and can easily lie outside the
physical temperature extremes that exist.  We conclude that
$T_{NIIwid/H\alpha wid}$ must be used with caution, unless one can be
sure that the nonthermal velocity is constant in the region. 

\subsection{Temperature Comparisons for Two Regions} \label{tworegions}

	Both [NII] and H$\alpha$ lines have been measured for two of our
objects, the $\lambda$ Ori HII region and BL. 

\subsubsection{$\lambda$Ori HII Region} \label{lambdao1}

	This HII region is excited by the star the O8III star $\lambda$
Ori; such a star has effective temperature 34000 K (Binney and
Merrifield (1998).  The radius of the He$^+$ ionization zone is about
0.35 the H$^+$ zone (Osterbrock 1989), so the He$^+$ volume is only
about 0.04 the H$^+$ volume; accordingly, we neglect the He$^+$ and take
$f_{He^+}=1$. 

	Our upper limit is $T_{4,R/H\alpha} < 0.71 \pm 0.05$.  The
exciting star has reddening $E(B-V)=0.12$ mag (Diplas and Savage 1994). 
If all of this extinction occurs in front of the HII region, then we
have $T_{4,R/H\alpha} = 0.43 \pm 0.03$. 

	Reynolds and Ogden (1982) measured the [NII] and H$\alpha$
lines; applying equation (11) of Haffner, Reynolds, and Tufte (1999)
gives $T_{4,NII/H\alpha} = 0.57 \pm 0.03$.  Also, Reynolds and Ogden
found $T_{4,NIIwid/H\alpha wid} = 0.60 \pm 0.17$ from a linewidth
comparison; if the uncertainty were smaller, then the agreement would
suggest that the temperature doesn't change much along the line of
sight.  We adopt the temperature from the intensity ratio,
$T_{4,NII/H\alpha} = 0.57 \pm 0.03$. 

	$T_{4,NII/H\alpha} - T_{4,R/H\alpha} = 0.14 \pm 0.04$.  The
difference is far larger than the uncertainty.  However, the results for
the toy model in Table~\ref{tmodels} suggest that we should expect a
difference of a few hundredths\footnote{The difference for the toy model
is 0.06, but the toy model considers only two extreme temperatures
instead of a continuous distribution and therefore overestimates the
difference.}.  Suppose that the true value of $T_{4,R/H\alpha}$ should be
a few hundredths smaller than $T_{4,NII/H\alpha}$, say 0.54. 

	There are two ways to make our derived $T_{4,R/H\alpha}$ equal
to 0.54.  For one way, the effective $E(B-V)$ is not 0.12 mag, but
instead the value required to reduce our upper limit of 0.71 to 0.54;
this would make the effective $E(B-V) = 0.06$ mag.  This could occur if
most of the extinction to the star $\lambda$ Ori occurs within the HII
region itself.  This is highly unlikely: the $\lambda$ Ori HII region
has a total column density to its center $N_e \sim 2 \times 10^{20}$
cm$^{-2}$ (Reich 1978), which corresponds to $E(B-V) \sim 0.03$ mag for
the normal gas/extinction ratio (Bohlin, Savage, and Drake 1978).  This
is much smaller than the total reddening to the star. 

	For the other way, we adjust the intensity scale errors to
attain agreement.  From equation~\ref{t4eqn} we require $\left({f_{2326}
\over f_{H\alpha}}\right)^{1.82} = {0.54 \pm 0.03 \over 0.43 \pm 0.03}$,
or ${f_{2326} \over f_{H\alpha}} = 1.13 \pm 0.06$. 

\subsubsection{BL-TOP} \label{bltop01}

	Again we take $f_{He^+} = 1$ because BL is far from the exciting
stars.  Our upper limit is $T_{4,R/H\alpha} < 0.49 \pm 0.07$.  There is
a previous radio determination by Gaylard (1984), who measured radio
recombination lines and continuum intensities at three positions; the
average of his temperatures is $T_{4,R/R} = 0.52 \pm 0.08$.  The
agreement between our upper limit and his value is good, which implies
that the reddening to BL-TOP is very low. 

	Reynolds and Ogden (1979) measured the [NII] and H$\alpha$
lines; again, applying equation (11) of Haffner, Reynolds, and Tufte
(1999) gives $T_{4,NII/H\alpha} = 0.61 \pm 0.03$.  Also, Reynolds and
Ogden found $T_{4,NIIwid/H\alpha wid} = 0.76 \pm 0.09$ from a linewidth
comparison.  These two temperatures differ by $2 \sigma$.  From our
discussion in Section~\ref{toymodel} above, we eliminate
$T_{4,NIIwid/H\alpha wid}$ from consideration and adopt the temperature
from the intensity ratio, $T_{4,NII/H\alpha} = 0.61 \pm 0.03$. 

	$T_{4,NII/H\alpha} - T_{4,R/H\alpha} = 0.12 \pm 0.08$.  The
difference is significant at the $1.5\sigma$ level.  However, it seems
that there is temperature structure within BL, so the expected
difference is perhaps $\sim 0.02$.  Thus, we ascribe the larger observed
difference to a scale error in $T_{4,R/H\alpha}$ and suppose that the
true value of $T_{4,R/H\alpha}$ should be $\sim 0.59$.  If there is no
extinction, then we must adjust the intensity scale errors to attain
agreement.  From equation~\ref{t4eqn} we require $\left({f_{2326} \over
f_{H\alpha}}\right)^{1.82} = {0.59 \pm 0.03 \over 0.49 \pm 0.07}$, or
${f_{2326} \over f_{H\alpha}} = 1.11 \pm 0.09$. 

\subsection{The Final Correction Factor} \label{finalcorrection}

	Above in Section~\ref{tworegions} we found two independent
values for the ratio of radio to optical correction factors.  They agree
well even though their uncertainties are rather large.  We adopt the
unweighted average ${f_{2326} \over f_{H\alpha}} = 1.12 \pm 0.07$; the
uncertainty does not include the errors in our guesses for the proper
differences $T_{4,NII/H\alpha} - T_{4,R/H\alpha}$. 

	We provisionally assign all of the calibration error to
$f_{H\alpha}$, for several reasons.  Firstly, the present H$\alpha$ data
are not corrected for sky transmission variations, either night-to-night
or airmass.  Secondly, the H$\alpha$ absolute calibration is ultimately
tied to the brightness of a $1^\circ$ diameter region in NGC7000 (the
North American Nebula), and the absolute intensity of this region is
uncertain at the level $\sim 12\%$.  Finally, Jonas (1999) has
determined the uncertainty in the 2326 MHz intensity to be $f_{2326} =
1.00 {+0.01 \over -0.05}$. 

	For the remainder of the present paper, we adopt $f_{2326}=1$
and $f_{H\alpha} = 0.89$. 

	For the ensuing discussion we do not use $T_{4,R/H\alpha}$.
Rather, we use values corrected by the factor $\left({f_{2326} \over
f_{H\alpha}}\right)^{1.82} = 1.23$ and denote these corrected values by
the symbol  $T_{4,R/H\alpha} \vert _{corr}$. These values are listed in
Table~\ref{barnardtable}. 

\section{ $T_{R/H\alpha}$: DISCUSSION FOR INDIVIDUAL OBJECTS}

\label{individualobjects}

	Equation~\ref{t4eqn} shows that the values for $T_{4,R/H\alpha}
\vert _{corr}$ listed in Table~\ref{barnardtable} are upper limits to
the electron temperature because the reddening $E(B-V)$ cannot be
negative. 

\subsection{The $\lambda$Ori HII Region}

	Following our discussion above in Section~\ref{lambdao1}, we
assume that all of the extinction occurs in front of the HII region and
adopt $E(B-V) = 0.12$.  For this choice, our upper limit becomes
$T_{4,R/H\alpha} \vert _{corr} = 0.53 \pm 0.04$.  By design, this is
consistent with $T_{4,NII/H\alpha} = 0.57 \pm 0.03$
(Section~\ref{absolutecal}). 

	This is considerably smaller than typical temperatures of HII
regions near the Sun.  In particular, the Ori A and Ori B HII regions
have measurements of $T_{4,R/R}$ ranging from $\sim 0.71 \rightarrow
0.86$ (Here the subscript $R/R$ means determined from the ratio of radio
recombination lines to radio continuum; see Reifenstein et al 1970,
Shaver et al 1983).  It is also smaller than the {\it predicted}
temperatures: for ``standard'' abundances, Osterbrock's (1989) Figure
3.2 should be reasonably representative of the $\lambda$ Ori HII region
and predicts an equilibrium temperature of $T_4 = 0.70$.  We now discuss
this discrepancy between observed and predicted temperatures for the
$\lambda$ Ori HII region. 

	The temperature of an HII region is governed primarily by the
abundances of Nitrogen and Oxygen.  Lowering the temperature from the
predicted $T_4 = 0.70$ to $T_4 = 0.57$ requires that the cooling be
increased by a factor of 1.3.  To attain the factor 1.3, the [OII]
and/or [NII] abundance must increase by the factor 1.8.  It is very
doubtful that this increase occurs because of an increased gas-phase
abundance produced by grain destruction, because Meyer et al (1997,
1998) find that only $\sim 30\%$ of the O and $\sim 0\%$ of the N are
locked up in grains. 

	An alternative to the observed temperature being as low as 5700
K is the fraction [N$^+$/N] being lower than unity, because the observed
value $T_{4,NII/H\alpha} = 0.57 \pm 0.03$ relies on the assumption that
N$^+$/N = H$^+$/H (Haffner et al 1999).  In fact, models by Sembach et
al (1999) suggest that this assumption is incorrect; instead, N$^+$/N =
0.7 H$^+$/H.  If this is correct, then the observed $T_{4,NII/H\alpha}$
becomes $0.62 \pm 0.03$.  This raises the temperature of the $\lambda$
Ori HII region and makes it closer to, but still different from, the
theoretical prediction. 

	We do not understand the causes for the smaller temperature of
the $\lambda$ Ori HII region. 

\subsection{BL-TOP}

	Following our discussion above in Section~\ref{bltop01}, we
adopt $E(B-V) = 0$.  This makes the temperature equal to our upper
limit, so $T_{4,R/H\alpha} \vert _{corr} = 0.61 \pm 0.09$.  Again, by
design this is close to $T_{4,NII/H\alpha} = 0.61 \pm 0.03$. 

\subsection{BL-LEFT}

	Our upper limit is $T_{4,R/H\alpha} \vert _{corr} < 1.06 \pm
0.18$.  This is considerably higher than BL-TOP.  We expect the physical
conditions in BL-LEFT and BL-TOP to be similar because they are part of
the same structure; this implies that the true temperature is lower than
our raw measurements imply and that extinction is important.  If BL-LEFT
has the same temperature as BL-TOP, then equation~\ref{t4eqn} predicts
$E(B-V) = 0.13$ mag. 

	This set of conditions is entirely in line with the ratio of
H$\alpha$ to H$\beta$ line intensities, which we denote by
$R_{\alpha\beta}$.  Isobe (1978) presents a map of this ratio from which
we estimate

\begin{equation}
\label{rab_barnard}
{ R_{\alpha\beta, BL-TOP} \over { R_{\alpha\beta, BL-LEFT} }} \sim 0.80 \ ,
\end{equation}

\noindent and a good approximation to the ratio of line intensities
(Martin 1988) is

\begin{equation}
\label{rab_theory}
 R_{\alpha\beta, theory} \approx 2.88 T_4^{-0.1} e^{1.05 E(B-V)} \ .
\end{equation}

\noindent These, together with the Table~\ref{barnardtable}'s ratio of
the slopes $B$ for the two regions, provide an independent estimate of
the quantities $T_{RATIO} = {T_{4,BL-TOP} \over T_{4,BL-LEFT} }$ and
$E_{DIFF} = [E(B-V)_{BL-TOP} - E(B-V)_{BL-LEFT}]$; we obtain $T_{RATIO}
\sim 0.58$ and $E_{DIFF} \sim -0.19$ mag, which agrees quite well with
our above results. 

	We conclude that BL-LEFT has the same temperature as BL-TOP,
$T_{4,R/H\alpha} \approx 0.61$.  Its reddening $E(B-V) \approx 0.13$ mag
corresponds to a foreground $N_H \sim 8.0 \times 10^{20}$ cm$^{-2}$.

\subsection{FILAMENT}

	Our upper limit is $T_{4,R/H\alpha} \vert _{corr} < 2.07 \pm
0.75$.  The uncertainty is large, but an eyeball examination of
Figure~\ref{halfa_radioplot} persuades us that this value is reliable
and is statistically significantly higher than in the other regions. 

	One spot in this region, $(\alpha, \delta) = (04^h00^m,
2.0^\circ)$, was observed by Reynolds and Ogden (1979, entry 3 in Table
1); these data give $T_{4,NII/H\alpha} = 0.55 \pm 0.03$.  Applying
equation~\ref{t4eqn}, we find $E(B-V) \sim 0.31$ mag.  This corresponds
to $N_H \sim 1.9 \times 10^{21}$ cm$^{-2}$, which in turn corresponds to
a 100 $\mu$m brightness of 28 MJy ster$^{-1}$ for the usual IR to HI
conversion of Schlegel, Finkbeiner, and Davis (1998).  The observed 100
$\mu$m brightnesses are smaller than this, varying from $\sim 6
\rightarrow 13$ MJy ster$^{-1}$.  This seeming discrepancy is not
meaningful because the HI column density is large and, moreover, the IR
emission per H-nucleus $\sim 0.7 \rightarrow 1.1$ MJy ster$^{-1}$
cm$_{20}^{-2}$, with the higher ratios correlated with the higher 100
$\mu$m brightnesses, which implies the presence of H$_2$.  Under these
dense conditions, the true column density exceeds that indicated by the
IR emission as discussed in section~\ref{dirbefits}. 

	We adopt the reddening required to make $T_{4,R/H\alpha}$ match
$T_{4,NII}$, 0.31 mag.  This weakens the observed brightness by a factor
$\sim 3.7$.  From Figure~\ref{halfa_radioplot}, we see that the observed
H$\alpha$ brightness is $EM_{H\alpha} \sim 90$ cm$^{-6}$ pc; thus, the
unabsorbed brightness $\sim EM_{H\alpha} \sim 340$ cm$^{-6}$ pc.  This
makes the intrinsic optical brightness of FILAMENT $\sim 0.6$ the
optical brightness of BL. 

\section{DIRBE FITS}

\label{dirbefits}

          IR emission traces the total warm/cold column density
[$N(Htot) = N(HI) + 2N(H{_2}) + N(H^+) $] and 21-cm line emission traces
the HI column density $N(HI)$; thus the appropriately-scaled difference
traces [$2N(H{_2}) + N(H^+)$].  For H$^+$, note the important difference
between its IR and H$\alpha$ (or radio) emission: IR traces column
density $N(H^+)$, while H$\alpha$ traces emission measure $EM = N(H^+)
n_e$.  This means that the comparison provides $n_e$.  

	The important scaling factor is $B$, the IR emission per
hydrogen nucleus.  The IR emission peaks near $100$ $\mu$.  Global fits
at 100 $\mu$m typically obtain $B$ in the vicinity of 0.65 MJy
ster$^{-1}$ cm$_{20}^{-2}$ (Schlegel et al 1998; Reach, Wall, and
Odegard 1998; Heiles, Haffner, and Reynolds 1999).  Here we adopt the
somewhat lower value 0.62 of Arendt et al (1998) because they obtained
the full DIRBE spectral coverage.  Let the emission per nucleus in
ionized gas be $b_{H^+}B$.  Then, neglecting H$_2$, we have

\begin{equation}
\label{idealfit1}
IR_{obs} = A + B N(HI) + b_{H^+}B N(H^+)  \; .
\end{equation}

          To treat observational data, consider a least-squares fit for
the the coefficients A, B, and C in the equation

\begin{equation}
\label{idealfit}
IR_{obs} = A + B N(HI) + C n_e N(H^+)  \; .
\end{equation}

\noindent Here $IR_{obs}$ is the DIRBE data, $N(HI)$ is the integrated
21-cm line intensity, and $N(H^+)n_e$ is from either the 2326 MHz data
or the velocity-integrated H$\alpha$ line intensity.  Comparing the last
term in these equations gives 

\begin{equation}
n_e = {b_{H^+} B \over C} \; .
\end{equation}

\noindent The radio data are better than the optical data for these fits
because they have higher angular resolution and are unaffected by
extinction; their higher noise is unimportant because BL is intense. 
The radio and optical data give comparable results.  In actual practice
we do not fit equation~\ref{idealfit} but instead

\begin{equation}
\label{realfit}
IR_{obs} = A + B N(HI) + C' EM_{2326}  \; .
\end{equation}

\noindent From equation~\ref{emradio} it is clear that

\begin{equation}
\label{neequation}
n_e = T_4^{0.35} f_{He^+}^{-1} f_{2326} b_{H^+} {B \over C'} = 0.84 b_{H^+}
{B \over C'} \; .
\end{equation}

\noindent where the numerical value is for $T_{4,R/H\alpha} \vert _{corr} =
0.61$, $f_{He^+} = 1$, and $f_{2326}=1$. 

	Reliable results require choosing appropriate regions,
specifically ones with small $N(HI)$ and no H$_2$.  High $N(HI)$ regions
are unsuitable for three reasons: (1) their possible saturation of the
21-cm line makes it an invalid tracer of $N(HI)$; (2) their associated
extinction shields the interior from starlight, making dust grains
cooler and reducing the IR emission per H atom, and (3) their probable
associated H$_2$ provides extra dust unrelated to HI, increasing the IR
emission per H atom. 

\subsection{Electron Density and Grain Emission Spectrum}

\label{irspectrum}

	Most of BL does not satisfy the low $N(HI)$ criterion because
there are very dense background molecular clouds.  Nevertheless, we were
able to locate two small portions of BL and its environs where molecular
clouds are absent.  These regions are outlined by dashed lines in
Figure~\ref{quad_diff}.  We performed fits at the six relevant DIRBE
wavelengths. 

	Figure~\ref{lsfitplot1} exhibits the logarithm of the IR
emission in MJy ster$^{-1}$ per $N(HI)_{20}$, where the subscript means
units of $10^{20}$ cm$^{-2}$; this is equal to $\log B$ in
equation~\ref{realfit}.  This figure also shows the global average
spectrum determined by Arendt et al (1998).  Our spectrum agrees well
with the global one. 

	At 100 $\mu$m, equation~\ref{neequation} gives ${n_e \over
b_{H^+}} = 0.84 \pm 0.15$ and $1.16 \pm 0.09$ cm$^{-3}$ for the top and
bottom regions, respectively.  These differ by $1.8\sigma$, which we
regard as fair agreement, and we adopt the average ${n_e \over b_{H^+}}
= 1.0$ cm$^{-3}$. 

	The ratio $C' \over B$ is the ratio of the H$^+$ and HI IR
spectra; if this ratio depends on wavelength, then the grain emission
spectrum differs for ionized and neutral gas.  Figure~\ref{lsfitplot}
exhibits this ratio as normalized to unity at $\lambda = 100$ $\mu$m. 
The H$^+$ spectrum differs significantly from the HI spectrum.  Relative
to $\lambda = 100$ $\mu$m, the 60 $\mu$m points are definitely larger,
and the 12 and 240 $\mu$m points are probably smaller (the errors are
large, particularly at 240 $\mu$m). 

	Our least-squares fit allows us to predict $IR_{pr} = A + B
N(HI) + C' EM_{2326}$.  Thus the difference $IR_{obs} - IR_{pr}$ (in
least-squares terminology, the residual) is an approximate measure of
$N(H_2)$ combined with the deviation of $n_e$ from its average, i.e. 

\begin{equation}
R = IR_{obs} - IR_{pr} = B \left[ 2 N(H_2) +
N(H^+) \left(1 - {n_e \over \langle n_e \rangle}\right) \right] \; .
\end{equation}

\noindent There is also a minor contribution from saturation of the
21-cm line.

          The lower right-hand quadrant in Figure~\ref{quad_diff} maps
this difference, which depends on two terms---the ``molecular'' and
``ionized'' terms.  The ionized term can be positive or negative: it is
negative in regions of large $n_e$, where a small column density gives
bright 2326 MHz emission.  Thus, dense HII regions would stand out as
deficiencies in Figure~\ref{quad_diff} if they were not associated with
molecular clouds.  However, the dominant contribution to this image is
clearly the Orion molecular clouds, which produce the large, prominent
burned-out bright blob.  

\subsection{The Geometrical Conundrum}

\label{conundrum}

	Above we adopted ${n_e \over b_{H^+}} = 1.0$ cm$^{-3}$.  The
2326 MHz brightness temperature $\sim 0.13$ K, which corresponds to
$N(H^+) n_e = 160$ cm$^{-6}$ pc.  Combining these with $b_{H^+} = 1$
gives $N(H^+)_{20} = 5.0$ and path length $L = 160$ pc. 

	This path length is very long: BL is at the distance of Orion,
about 450 pc, so our 160 pc line-of-sight length is $36\%$ of the total
distance.  This, in turn, implies that BL is a cylinder that happens to
have its axis accurately pointed towards us.  This is not only deeply
unsatisfying, but also at odds with what we would expect from the simple
geometrical and physical model of a thick-walled spherical shell. 

	In the vicinity of the fitted regions, BL has radius $R \sim
6.1^\circ$ and apparent thickness $T \sim 1.3^\circ$; these translate to
lengths $R = 48$ and $T = 10$ pc.  Such a shell has a maximum tangential
path length through its edge of $L_{sph} \sim 60$ pc.  Our derived path
length exceeds this by a factor of 2.7; not only that, it exceeds even
the diameter by a factor of 1.6! This discrepancy is the geometrical
conundrum. 

	Clumping makes the conundrum even worse.  Suppose that the line
of sight within BL contains clumps that occupy a fraction $f$ of the
sightline.  This does not affect any of our derived physical parameters,
including $L$; however in this case $L$ refers to the length of the line
of sight {\it within the clumps}.  The {\it total} length within which
the clumps lie is $L/f$.  This makes the geometrical conundrum worse by
a factor of $f$. 

	To resolve the conundrum comfortably, we need to decrease $L$
to something less than its maximum of 60 pc. For discussion purposes, we
will assume $L = 40$ pc. This means we need to decrease our derived path
length by a factor of 4.   

	Suppose that ionized regions exhibit more grain emission per
$N(H_{tot})$ than neutral regions; this means $b_{H^+} >1$.  This
increases $n_e$ by the factor $b_{H^+}$ and decreases $L$ by the {\it
square} of $b_{H^+}$.  To resolve the conundrum in a comfortable way, we
need $b_{H^+}^2 \sim 4$ or $b_{H^+} \sim 2$. 

	We adopt this as the resolution to the conundrum.  Thus, we
adopt $n_e = 2.0$ cm$^{-3}$.  BL has $T_{4,R/H\alpha} \vert _{corr}
\approx 0.61$; thus, $ \tilde P_4 \equiv {P \over 10^4 k} = 2 n_e T_4
\approx 2.4$. 

\section{THE BL BIG GRAINS ARE WARMER THAN HI BIG GRAINS}

\label{warmgrains}

	Above in Section~\ref{conundrum}, we found the geometrical
conundrum requires more 100 $\mu$m IR emission per grain than in HI
regions, and we adopted $b_{H^+} = 2$.  And in Section~\ref{irspectrum},
we found the IR spectrum of BL to be different from that of HI.  Both
the increased 100 $\mu$m emission and the changed spectrum suggest that
the BL grains are warmer than the HI grains.  In the remainder of this
section, we will explore this temperature increase.  We shall assume
that the {\it number} of 100 $\mu$m emitting grains per H-nucleus is
identical in BL and HI regions. 

	We perform least square fits of the IR spectrum to determine the
big grain temperature $T_{BG}$.  In all fits we use a $\nu^2$ emissivity
law and weight the points by the inverse square of the uncertainties. 
In HI regions, the $\lambda \gtrsim 100$ $\mu$m radiation comes almost
exclusively from big grains ({\it BG}) and 60 $\mu$m radiation from very
small grains ({\it VSG}; D\'esert, Boulanger, and Puget 1990).  These
are two distinct grain populations and in solving for the temperature of
$\lambda = 100$ $\mu$m emitting grains we must consider only the BG's,
i.e.~we must use IR intensities for $\lambda \ge 100$ $\mu$m.  For the
three HI grain spectra in Figure~\ref{lsfitplot1} we obtain $T_{BG} =
(18.2, 15.4, 17.5)$ K for the (global, top, and bottom) regions,
respectively.  These refer to HI regions.  The uncertainties on our
derived temperatures (top, bottom) are large, so below we will use the
global $T_{BG} = 18.2$ K as a reference. 

	In BL, Figure~\ref{lsfitplot} shows 60 $\mu$m is much stronger
than in HI regions by a factor of 1.9 to 3.4.  This can occur for two
reasons.  One, the population of VSG's is enhanced in BL relative to HI
regions.  Two, the temperature of BG's is considerably larger in BL than
in HI regions. 

	Below, we will least-squares fit the grain temperatures to the
IR spectra for the H$^+$ in BL (top, bottom) and exhibit the results in
Table~\ref{graintable}.  The table also displays two important related
quantities.  One is $ b_{H^+}$, the ratio of the 100 $\mu$m emissivity
at the derived $T_{BG}$ in BL to that in the global HI.  The other is
$\left( T_{BG} \over T_{BG,HI} \right)^6$: for a grain emissivity
$\propto \nu^2$, this is the expected ratio of grain heating (equal to
total grain IR emission). 

	To begin, we suppose that the enhanced 60 $\mu$m intensity in BL
comes from BG's that are warm enough to overwhelm the VSG emission. 
Therefore, in this fit we derive $T_{BG}$ by including the amount of 60
$\mu$m intensity that exceeds what we expect from the the VSG's,
i.e.~the 60 $\mu$m emission above the dotted line in
Figure~\ref{lsfitplot}.  For H$^+$ in BL (top, bottom) we obtain $T_{BG}
= (24.2, 22.4)$ K.  These correspond to $b_{H^+} = (6.9, 4.2)$, which is
much larger than the $b_{H^+} = 2$ estimate above.  This can be
reconciled by making the clumping factor $f \neq 1$.  If the physical
path length over which the H$^+$ is distributed is fixed at 40 pc, then
$f \propto b_{H^+}^{-2}$; if $b_{H^+} = 2$, as we argue above, then we
require $f = \left[{(6.9, 4.2) \over 2}\right]^2 = (11.9, 4.4)$ and the
path lengths actually occupied by the H$^+$ in BL are $L \sim (4, 9)$
pc.  This is a high degree of clumping, and we would expect to see
considerable small-scale structure of the H$\alpha$ emission within BL. 
However, the image presented by Isobe (1973) doesn't give this
impression.  We conclude that the enhanced 60 $\mu$m intensity in BL
does {\it not} come from warm BG's. 

	We next consider the other alternative, namely that the enhanced
60 $\mu$m comes from an enhanced population of VSG's. In this fit, 
we derive $T_{BG}$ in the usual way, by including only the $\lambda \ge 100$
$\mu$m data. For H$^+$ in BL (top, bottom) we obtain $T_{BG}
= (24.2, 22.4)$ K.  These correspond to $b_{H^+} = (1.7, 2.2)$, which
agrees well with the $b_{H^+} = 2$ estimate above.  These grains require
an excess heating rate relative to global HI by factors of $(1.5, 1.9)$.

	This excess heating rate is very close to that expected from the
added heating by L$\alpha$ photons trapped in ionized gas.  Spitzer
(1978) shows that this mean intensity of these photons is $I_{L\alpha}
\approx 1.1 \times 10^{-3} n_e$ erg cm$^{-2}$ s$^{-1}$ str$^{-1}$.  The
ISRF is equivalent to a blackbody at 3.14 K (Mathis, Mezger, and Panagia
1983), or $I_{ISRF} \approx 1.8 \times 10^{-3}$ erg cm$^{-2}$ s$^{-1}$
str$^{-1}$.  For $n_e = 2$ cm$^{-2}$ (Section~\ref{conundrum}), we have
${I_{ISRF} + I_{L\alpha} \over I_{ISRF}} = 2.3$, which is close to the
excess heating obtained in the above paragraph. 

	We emphasize that this extra grain heating is a
generally-occuring process in all $H^+$ regions that trap L$\alpha$
photons (the ``on-the-spot'' case).  The H$^+$ grain heating rate
$\approx (1 + 0.6 n_e)$ times larger than the HI grain heating rate. 
This translates into an increased IR emission per grain.

	We much prefer this second alternative because the it is
consistent with the $b_{H^+} = 2$ estimate obtained from our geometrical
argument in Section~\ref{conundrum}.  Also, the increase in $T_{BG}$ is
modest and is attained with something close to the expected increase in
grain heating rate.  We conclude that the 60 $\mu$m excess in the H$^+$
of BL comes from an enhanced population of VSG's. 

	In contrast, the relative 12 $\mu$m intensity from BL is smaller
than that from the global HI.  This result is strongly suggested from
Figure~\ref{lsfitplot} but, given the uncertainties, not absolutely
certain.  The 12 $\mu$m emission comes from a third population of
grains; many workers, including D\'esert et al (1990), believe they are
PAH's.  We conclude that, with high probability, the population of PAH's
relative to BG's in the H$^+$ gas of BL is smaller than in global HI. 

	Our conclusions, then, are that the VSG's are more abundant and
the PAH's less abundant in BL than in global HI.  This is similar to but
firmer than the results of D\'esert et al (1990) on the H$^+$ gas in the
California nebula.  Clearly, it would be desirable to confirm this
trend by studying other HII regions. 

\section{CONCLUSIONS}

\label{conclusions}

	In Section~\ref{barnardradiosection}, we derived the radio
spectral index of the Barnard Loop (BL) from large-scale radio surveys
at four frequencies.  We eliminated possible scale errors by assuming
the HII region $\lambda$ Ori to be a thermal (optically thin
bremstrahlung) radio source.  We found BL to have a thermal spectrum,
too. 

	Having found the radio emission of BL to be thermal, we could
combine the radio and H$\alpha$ line data in
Section~\ref{barnardrjrsection} to derive upper limits on the electron
temperature (Table~\ref{barnardtable}) for four regions: the $\lambda$
Ori HII region, two regions in BL, and a high-latitude filament in the
wall of the Orion-Eridanus superbubble.  In Section~\ref{absolutecal} we
discussed two of these regions in detail and compared our temperatures
with those previously obtained from NII and H$\alpha$ line ratios.  Our
currently-derived temperatures were lower than the previous ones.  We
developed a toy model to explore the effect of temperature structure
along the line of sight, and found that one expects the radio/H$\alpha$
temperatures to be somewhat smaller than the NII/H$\alpha$ temperatures. 
However, most of the discrepancy is a result of inaccurate absolute
intensity calilbrations.  Provisionally, we accepted the 2326 MHz radio
survey calibration as accurate (Jonas et al 1998) and ascribed all the
error to the WHAM survey data, whose final calibration has not yet been
done; the current WHAM intensities need to be multiplied by the factor
0.89. 

	We discussed the four regions in some detail and adopted
reddenings, which enabled us to derive actual electron temperatures
(Table~\ref{barnardtable}).  Temperatures in all four regions are
somewhat smaller than in the Orion nebula and other HII regions near the
Sun. 

	In Section~\ref{dirbefits} we performed least-squares fits of
the DIRBE diffuse IR intensities to the 21-cm line and radio continuum
intensities.  Our derived IR spectrum spans $\lambda = 12 \rightarrow
240$ $\mu$m and agrees well with the spectrum from global HI derived by
Arendt et al (1998).  In this fit, the ratio of the derived coefficients
of the 21-cm line and radio continuum allow one to determine the
electron density $n_e$.  However, our derived $n_e$ is small, requiring
a very long path length through the H$\alpha$ emitting region, which is
inconsistent with the BL morphology and leads to the ``geometrical
conundrum'' of Section~\ref{conundrum}. 

	The conundrum can be resolved if the H$^+$ grains emit more IR
per H-nucleus than HI grains do.  This, together with the modified IR
spectrum from the H$^+$ gas, is explained in Section~\ref{warmgrains} by
a higher grain temperature in the H$^+$ than in the HI ($T_{BG}$ in
Table~\ref{graintable}: HI is the first entry, H$^+$ the last two
entries).  The increased grain temperature in the H$^+$ gas agrees well
with that expected from the extra heating produced by trapped L$\alpha$
photons. The H$^+$ grain heating rate is generally higher than the HI
heating rate by a factor $\approx (1 + 0.6n_e)$. 

	In BL, $n_e \approx 2.0$ cm$^{-3}$ and $T \approx 6100$ K.  The
pressure is ${P \over k} \approx 24000$ cm$^{-3}$ K. In addition,
the H$^+$ grains in BL exhibit excess 60 $\mu$m emission and deficient
12 $\mu$m emission, indicating that very small grains (VSG's) are more
abundant and PAH's less abundant in BL than in the global HI.

\acknowledgements

	We thank D.~Finkbeiner for much pleasurable and instructive
consultation and for supplying destriped versions of the IR and 408 MHz
datasets in a convenient form; J.~Jonas for a copy of his thesis, the
2326 MHz dataset, and instructive discussions; P.~and W.~Reich for
instruction and consultation on the three low-frequency radio continuum
datasets; and C.~Salter for discussions on the radio properties of BL. 
WHAM is supported by National Science Foundation grant AST9619424, and
CH is supported in part by grant AST9530590.

\clearpage

\figcaption{Three scaled difference images, with thermal radiation and an
arbitrary zerolevel subtracted away; and the 2326 MHz image, as
labeled.  Each image is about $26^\circ$ in size and $b$ increases
upwards.  The two black circles are the Ori A and Ori B HII regions. 
The dotted line guides the eye to BL; the dashed lines enclose ON and
OFF regions for the $\lambda$Oph HII region.  The rectangles aid the eye
in interpreting the greyscale calibration (see text). \label{quadradio}}

\figcaption{Two pairs of 2326 MHz radio and H$\alpha$ optical images
side-by-side; each image is about $26^\circ$ in size and $b$ increases
upwards.  The upper set exhibits BL and the $\lambda$Ori HII region; the
two burned-out blobs are Ori A and Ori B.  The bottom set exhibits the
FILAMENT, which lies near $b = -40^\circ$.  The optical 
data integrate over a $1^\circ$-diameter circle and
are spaced by about $1^\circ$; the positions observed are marked by
dots.  The dashed lines enclose regions where least squares fits were
done (see text). \label{quadoptical}}

\figcaption{Data points and least squares fits for the four regions shown
on Figure~\ref{quadoptical}. \label{halfa_radioplot}}

\figcaption{Three images of data used for the DIRBE least-squares fits. 
The dashed lines outline the small regions where background molecular
clouds are absent and allow a reasonably accurate least squares fit of
equation~\ref{realfit}.  Each image is about $26^\circ$ in size and $b$
increases upwards. \label{quad_diff}}

\figcaption{$B$ is the logarithm of the IR brightness (MJy ster$^{-1}$) per
$10^{20}$ H-nuclei in the neutral gas; the arrow marks the global
average.  Dashed and solid lines are for the top and bottom regions in
Figure~\ref{quad_diff}, respectively; the dotted line and global average
are from Arendt et al (1998). \label{lsfitplot1} }

\figcaption{$C' \over B$, which is the ratio of the H$^+$ and HI IR
spectra, normalized to unity at $\lambda = 100$ $\mu$m.  Dashed and
solid lines are for the top and bottom regions in
Figure~\ref{quad_diff}, respectively; the dotted line is the global
average of Arendt et al (1998).  If we had derived $n_e$ at wavelengths
other than 100 $\mu$m, the derived $n_e$ would equal 
$B \over C'$ times the 100 $\mu$m value. \label{lsfitplot}}

\clearpage

\begin{deluxetable} {crrrrrr} 
\footnotesize
\tablecaption{ELECTRON TEMPERATURES \label{barnardtable}}
\tablewidth{450pt}
\tablehead{
\colhead{REGION} & \colhead{A} & \colhead{B} & 
\colhead{${f_{He^+}^{1.82} T_{4,R/H\alpha} \over e^{-4.25 E(B-V)}}$} &
\colhead{$T_{4,R/H\alpha} \vert _{corr}$} & $E(B-V)$ & 
\colhead{$T_{4,R/H\alpha} \vert _{corr}$} 
\nl 
& & & & \colhead{(upper limit)} & \colhead{(adopted)} & 
\colhead{(adopted)}
}
\startdata
$\lambda$Ori  & $ 371 \pm 7  $ & $ 0.83 \pm 0.03 $ & $0.71 \pm 0.05$ 
                   & $ <0.88 \pm 0.06$ & 0.12 & $0.53$ \nl 
BL-TOP        & $ 350 \pm 15 $ & $ 0.67 \pm 0.05 $ & $0.49 \pm 0.07$ 
                   & $ <0.61 \pm 0.09$ & 0 & $0.61$ \nl 
BL-LEFT       & $ 204 \pm 17 $ & $ 0.83 \pm 0.14 $ & $0.86 \pm 0.14$ 
                   & $ <1.06 \pm 0.18$ & 0.13 & $0.61$ \nl 
FILAMENT      & $ 161 \pm 13 $ & $ 1.63 \pm 0.59 $ & $1.68 \pm 0.61$ 
                   & $ <2.07 \pm 0.75$ & 0.31 & $0.55$ \nl 
\enddata 
\tablecomments{A and B are results of the least square fits of
$EM_{2326}$ to $EM_{H\alpha}$ (Equation~\ref{radoptlsfit} and
Figure~\ref{halfa_radioplot}). 
$T_{4,R/H\alpha} \vert _{corr}$ in column 5 is identical to column 4,
but corrected for intensity scale errors (Section~\ref{absolutecal}); it
is an upper limit.  Column 6 is the adopted reddening.  Column 7 is the
actual value of $T_{4,R/H\alpha} \vert _{corr}$ using the adopted
reddening in Column 6.  } \end{deluxetable}

\begin{deluxetable} {cccrcc} 
\footnotesize
\tablecaption{TEMPERATURES DERIVED FROM A TWO-COMPONENT TOY MODEL 
\label{tmodels}}
\tablewidth{450pt}
\tablehead{
\colhead{VTURB, RGN 1} & \colhead{VTURB, RGN 2} & 
\colhead{$T_{4,NIIwid/H\alpha wid}$} &
\colhead{VTURB} &
\colhead{H$\alpha$ WIDTH}  &
\colhead{NII WIDTH}  
}
\startdata
17  & 17     & 0.547 & 17.19 & 26.88 & 21.39  \nl
 7  &  7     & 0.542 &  7.47 & 21.91 & 14.75  \nl
12  &  7     & 0.678 &  7.86 & 23.43 & 15.10  \nl
17  &  7     & 0.872 &  7.96 & 25.28 & 15.36  \nl
7   & 12     & 0.408 & 11.72 & 22.37 & 17.17  \nl
7   & 17     & 0.210 & 15.85 & 22.88 & 20.05  \nl
\enddata

\tablecomments{We assume regions 1 and 2 have equal emission measures with
$T_4=0.45$ and 0.80, respectively; and ${\rm \left({N \over H}\right)} =
5.43 \times 10^{-5}$.  For all cases, ${I_{NII} \over I_{H\alpha}} =
0.229$, $T_{R/H\alpha} = 0.570$, and $T_{NII/H\alpha} = 0.630$. Columns 
3 and 4 are the temperature and nonthermal
linewidth that would be derived from the observed NII and H$\alpha$
lines.  All
velocities are halfwidths in km s$^{-1}$.}

\end{deluxetable}

\begin{deluxetable} {cccc} 
\footnotesize 
\tablecaption{GRAIN
TEMPERATURES, 100 $\mu$m EMISSIVITY, AND HEATING RATES \label{graintable}} 
\tablewidth{350pt} 
\tablehead{ \colhead{CONDITION} &
\colhead{$T_{BG}$, K} & 
\colhead {$b_{H^+}$ } & 
\colhead{$\left( T_{BG} \over T_{G,HI} \right)^6$} 
} 
\startdata 
                GLOBAL HI & 18.3 & 1.0 & 1.0 \nl 
H$^+$ with 60 $\mu$m, TOP & 24.2 & 6.9 & 5.4 \nl
H$^+$ with 60 $\mu$m, BOT & 22.4 & 4.2 & 3.4 \nl
 H$^+$ w/o 60 $\mu$m,  TOP & 19.6 & 1.7 & 1.5 \nl
 H$^+$ w/o 60 $\mu$m, BOT & 20.3 & 2.2 & 1.9
\nl

\enddata \tablecomments{GLOBAL HI temperatures are from our fit to
the spectrum of Arendt et al (1998); all temperature fits assume a
$\nu^2$ emissivity law, so Column 4 is proportional to the grain 
heating rate.  In Column 3, $b_{H^+}$ is the $\lambda
= 100$ $\mu$m grain emissivity in the H$^+$ relative to that in HI.  }
\end{deluxetable}

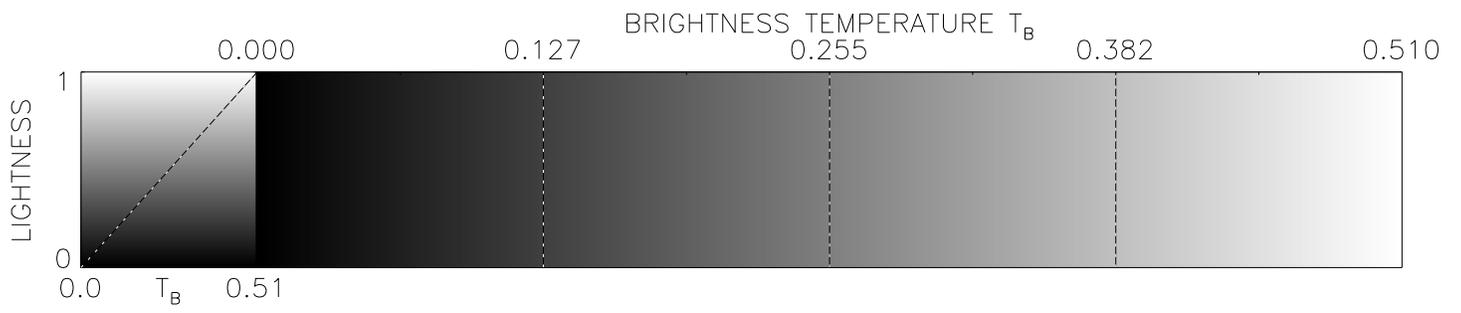
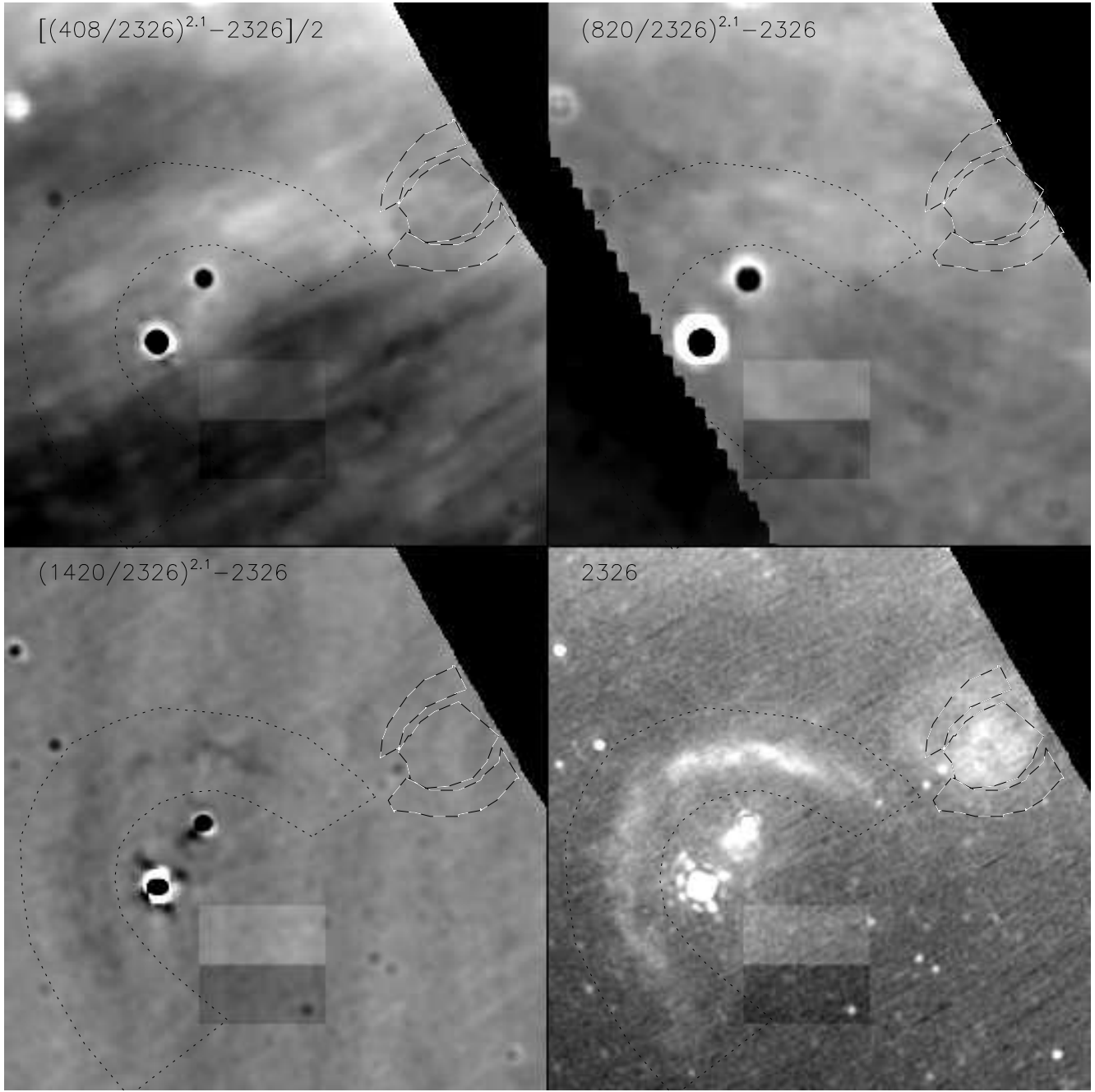
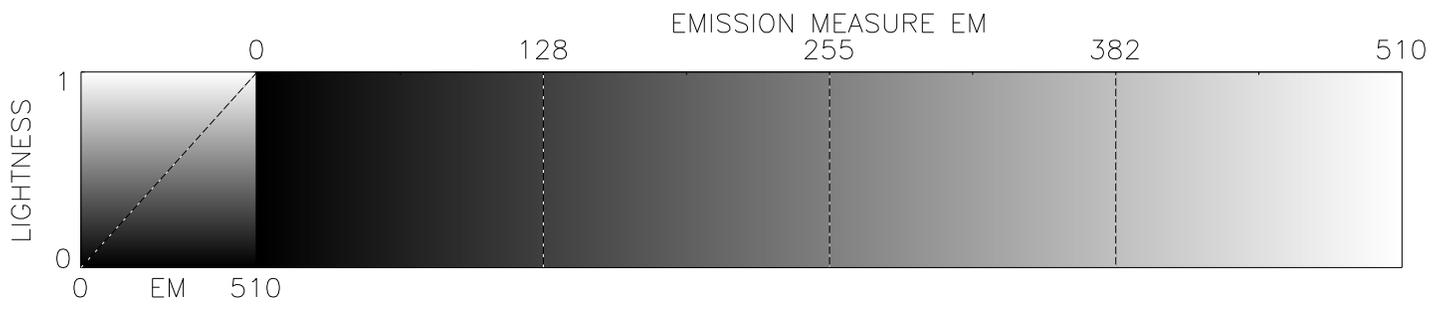
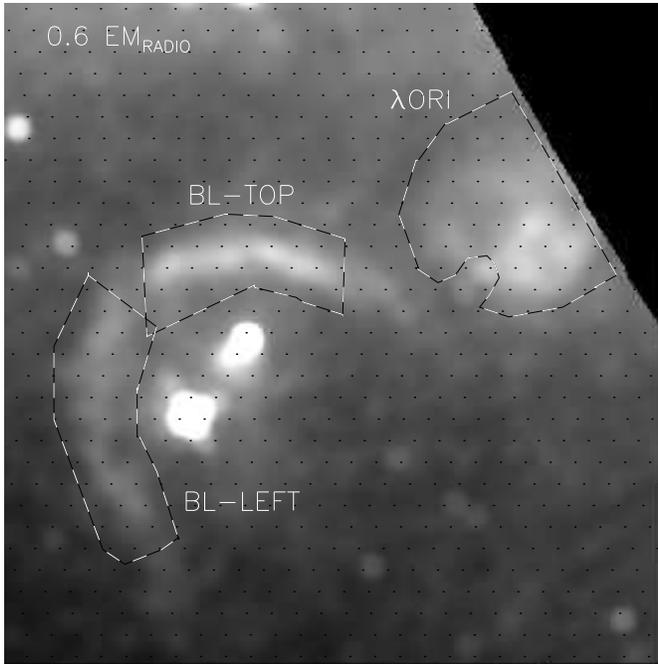
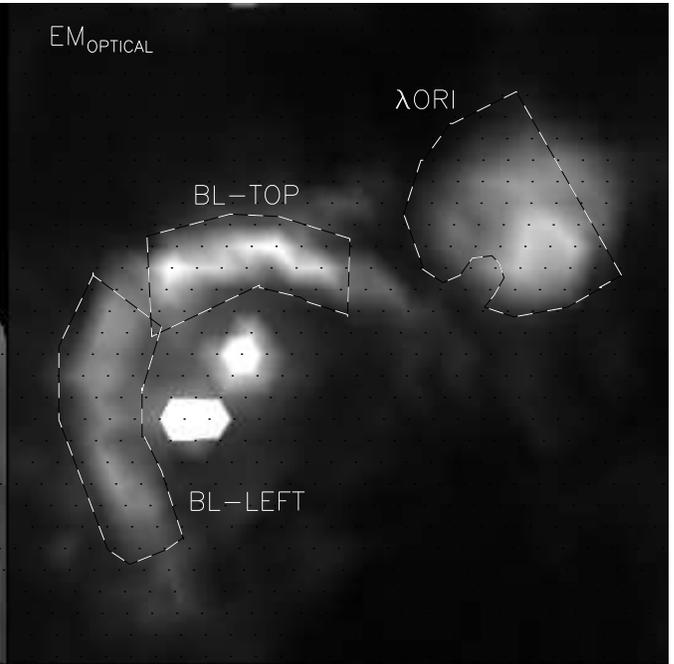
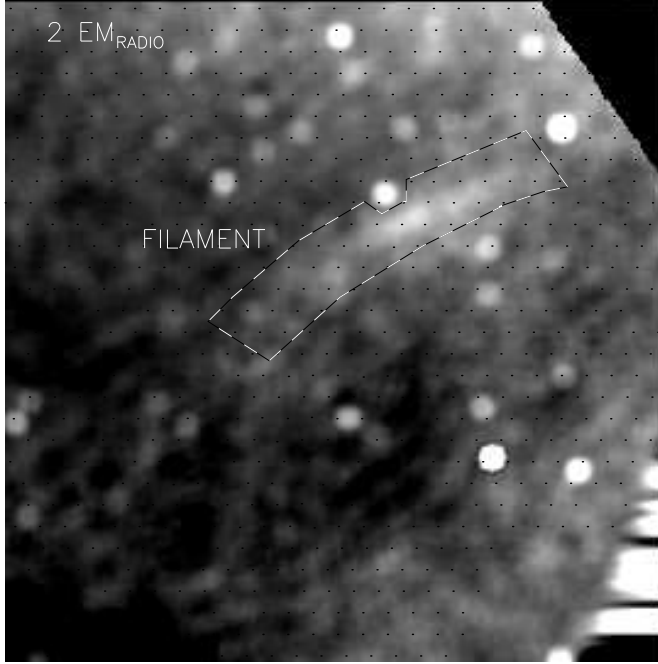
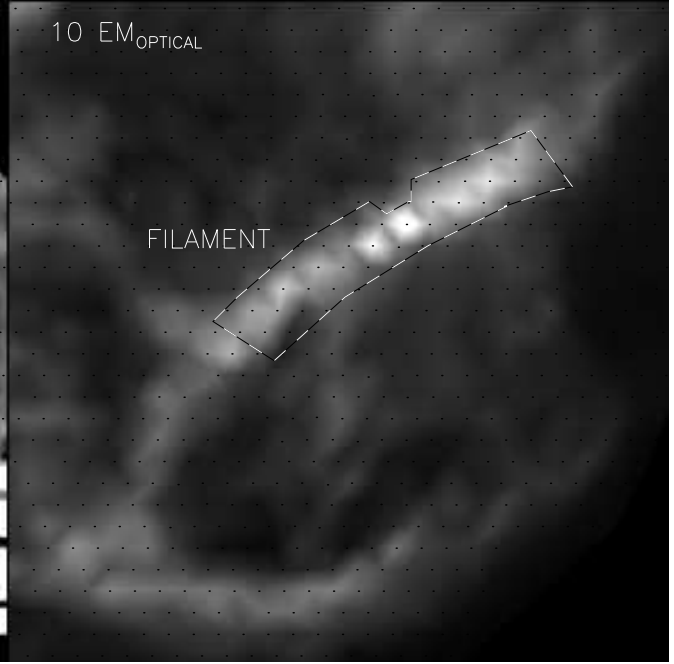
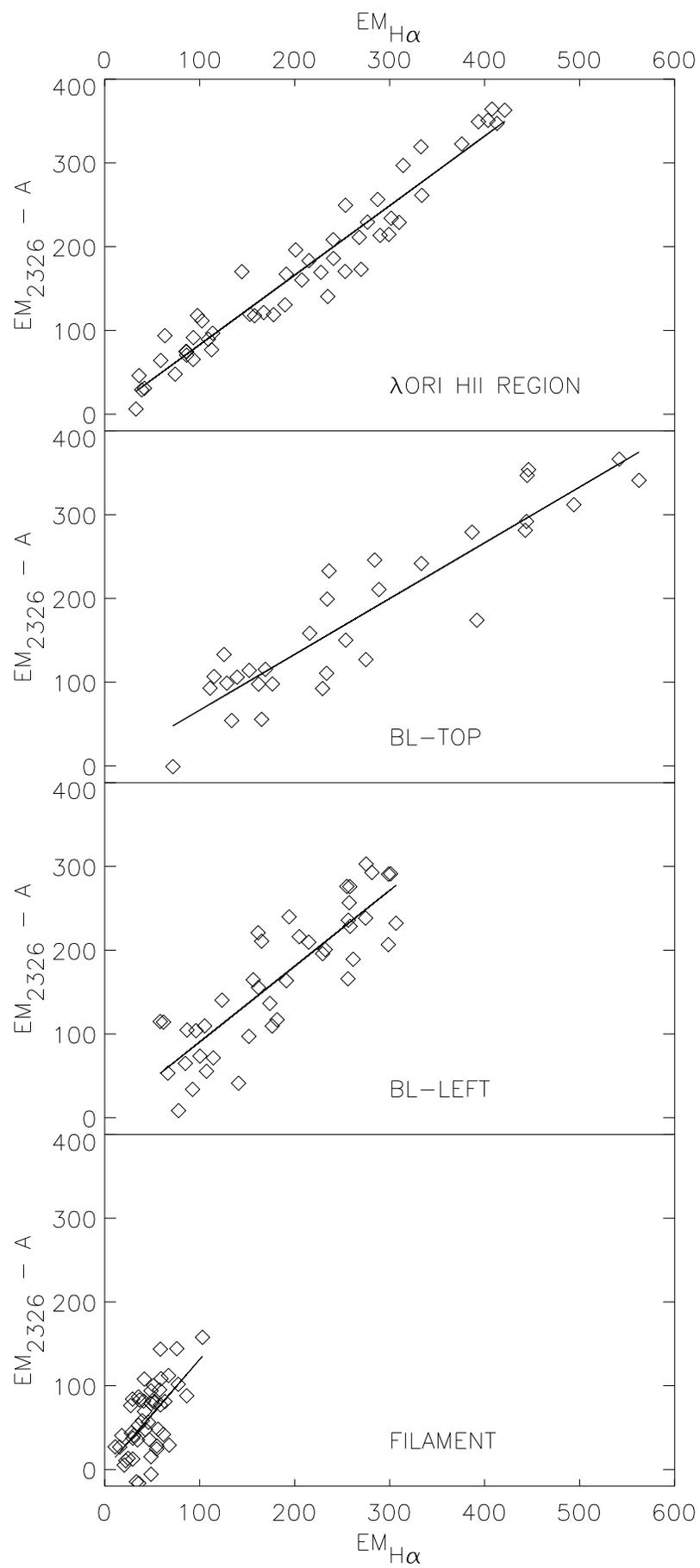
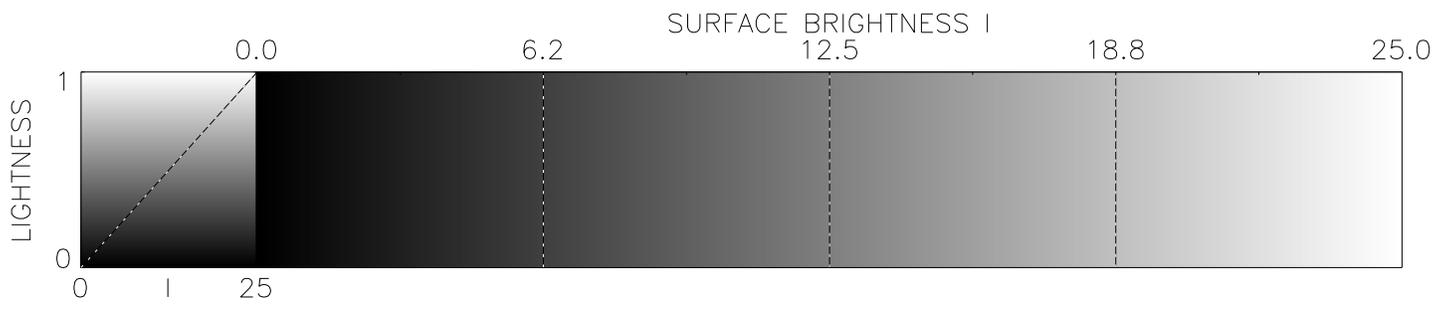
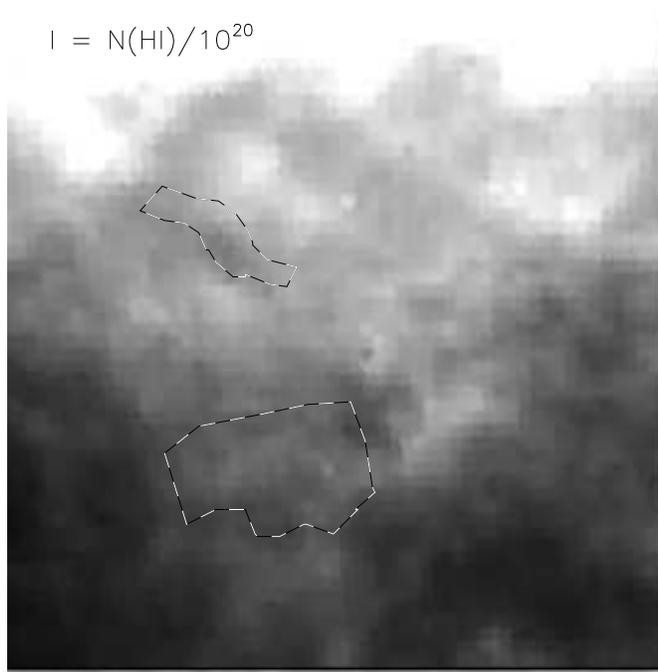
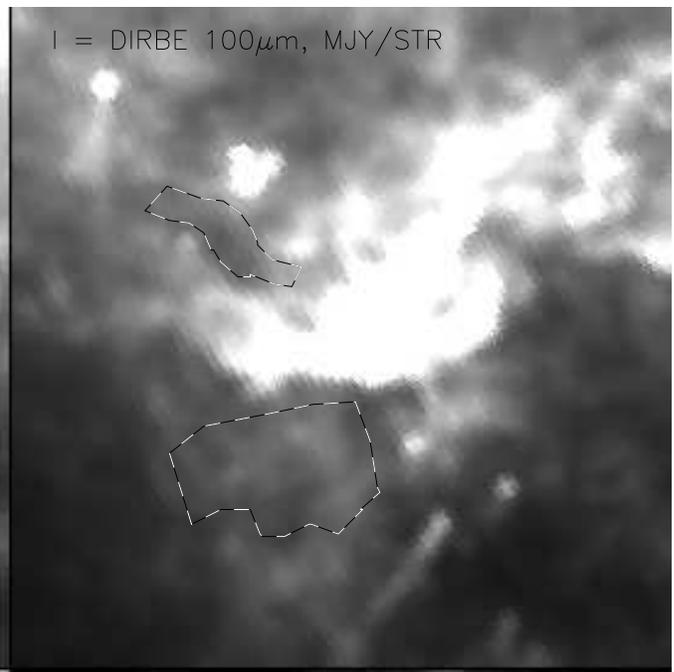
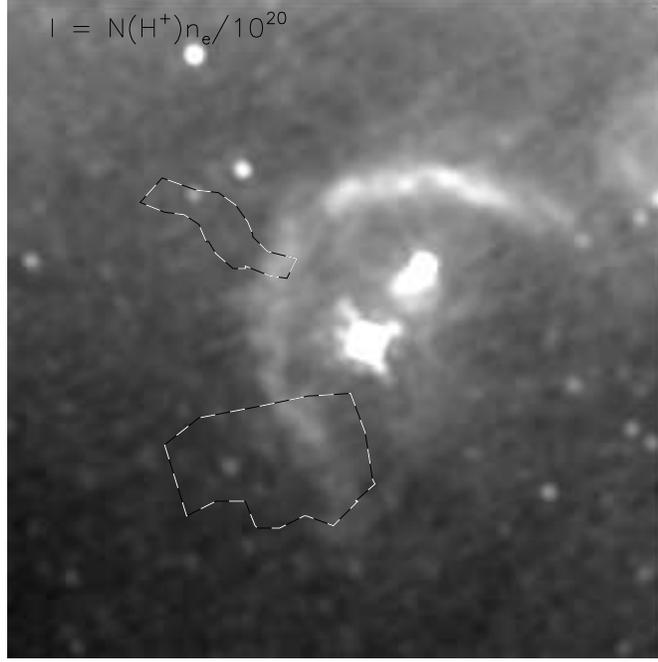
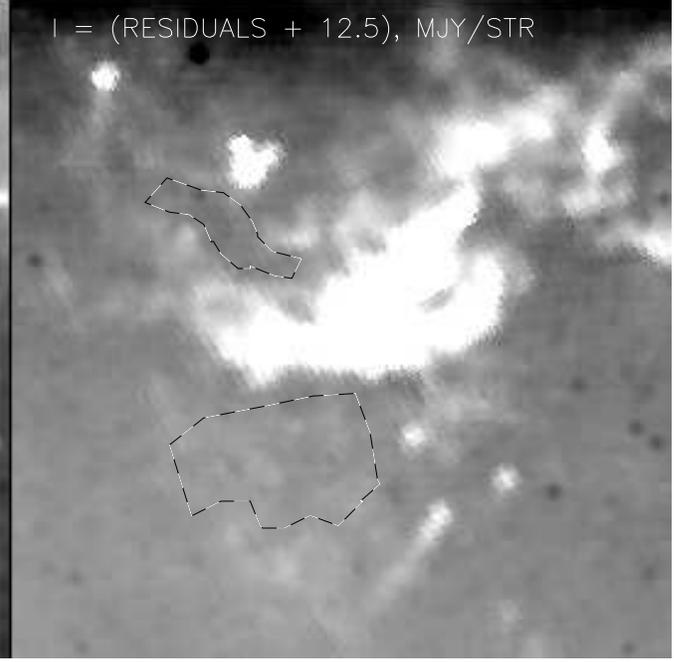
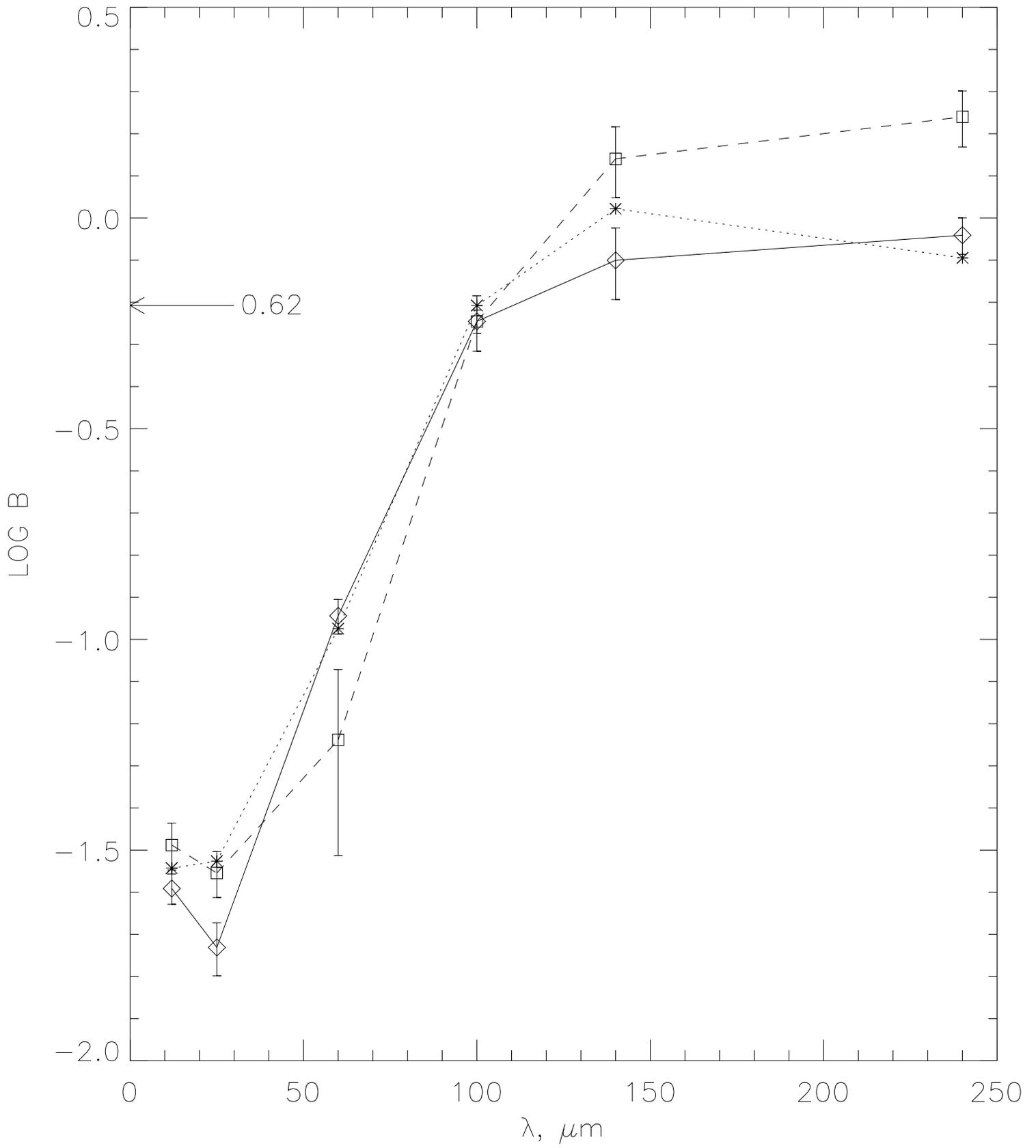
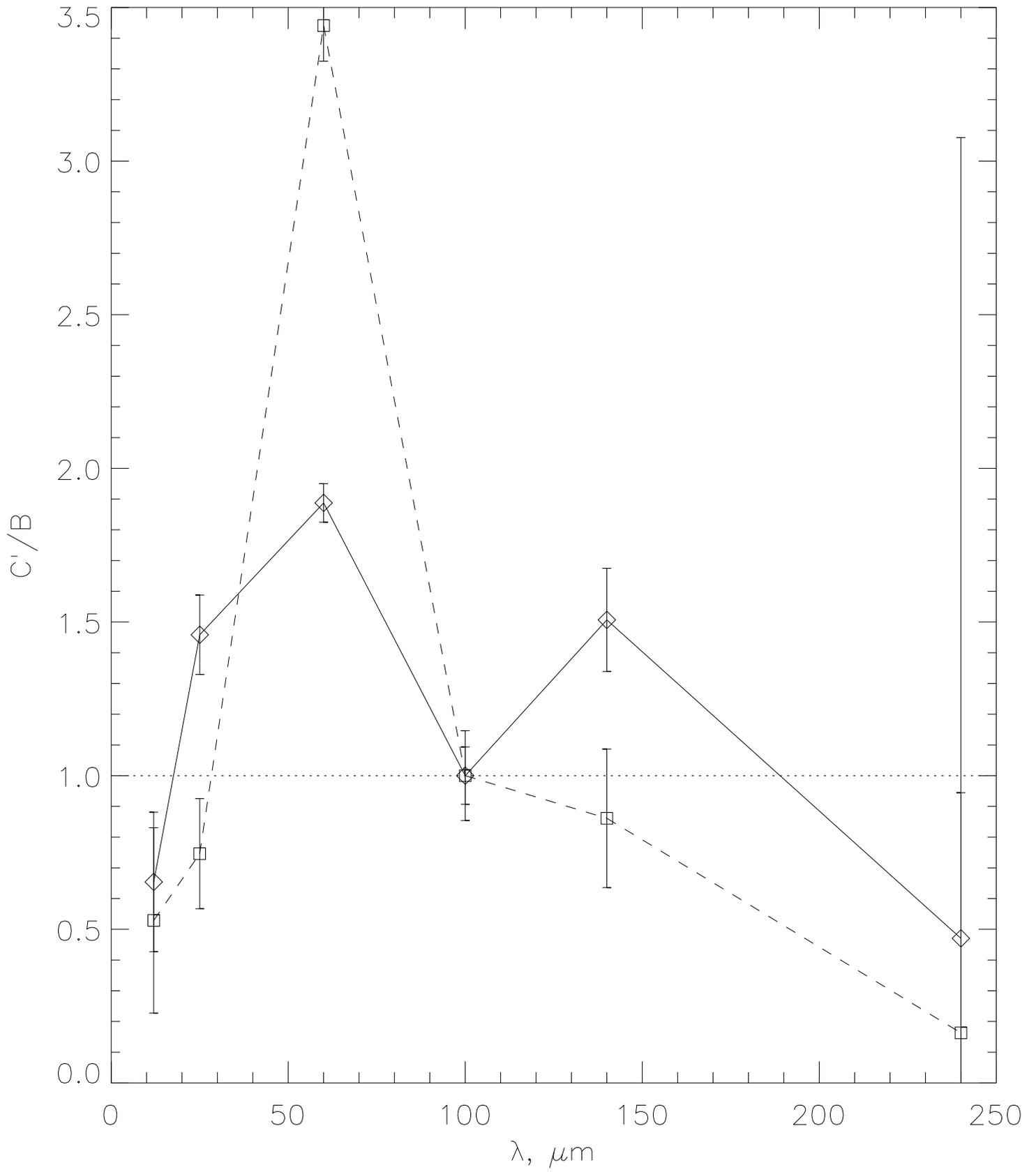
\end{document}